\documentclass[12pt, preprint]{aastex}
\usepackage{emulateapj5}
\usepackage{onecolfloat5}
\usepackage{apjfonts}
\usepackage{epsfig}

\def\myputfigure#1#2#3#4#5%
{\vskip#5pt\makebox[0pt]{\hskip#2in
\includegraphics[width=#3\textwidth]{#1}}\vskip#4pt\hfill}

\newcommand\lsim{\mathrel{\rlap{\lower4pt\hbox{\hskip1pt$\sim$}}
        \raise1pt\hbox{$<$}}}
\newcommand\gsim{\mathrel{\rlap{\lower4pt\hbox{\hskip1pt$\sim$}}
        \raise1pt\hbox{$>$}}}

\newcommand{\lya}{Lyman~$\alpha~$}

\newcommand{\del}{\delta(M)}

\newcommand{\delc}{\delta_c(M, z)}
\newcommand{\sig}{\sigma(M)}
\newcommand{\sigsq}{\sigma^2(M)}
\newcommand{\sfrd}{\dot{\rho}_\ast(z)}
\newcommand{\GRBrd}{\dot{\rho}_{\rm GRB}(z)}

\newcommand\msun{\rm M_\odot}
\def\gsim{\;\rlap{\lower 2.5pt
 \hbox{$\sim$}}\raise 1.5pt\hbox{$>$}\;}
\def\lsim{\;\rlap{\lower 2.5pt
   \hbox{$\sim$}}\raise 1.5pt\hbox{$<$}\;}

\begin{document}
\twocolumn[
\title{Constraints on the Small-Scale Power Spectrum of Density Fluctuations\\ From High-Redshift Gamma-Ray Bursts}

\author{Andrei Mesinger}
\affil{Department of Astronomy, Columbia University, 550 West 120th Street, New York, NY 10027}

\author{Rosalba Perna}
\affil{Department of Astrophysical and Planetary Sciences, University of Colorado, Boulder, CO 80309-0391}

\and
\vspace{-0.4cm}
\author{Zolt\'{a}n Haiman}
\affil{Department of Astronomy, Columbia University, 550 West 120th Street, New York, NY 10027}
\vspace{+0.4cm}

\submitted{Accepted for publication in ApJ}

\begin{abstract}
Cosmological models that include suppression of the power spectrum of
density fluctuations on small scales exhibit an exponential reduction
of high--redshift, non--linear structures, including a reduction in
the rate of gamma ray bursts (GRBs).  Here we quantify the constraints
that the detection of distant GRBs would place on structure formation
models with reduced small--scale power.  We compute the number of GRBs
that could be detectable by the {\it Swift} satellite at high
redshifts ($z\gsim 6$), assuming that the GRBs trace the cosmic star
formation history, which itself traces the formation of non--linear
structures. We calibrate simple models of the intrinsic luminosity
function of the bursts to the number and flux distribution of GRBs
observed by the {\it Burst And Transient Source Experiment
(BATSE)}. We find that a discovery of high--$z$ GRBs would imply
strong constraints on models with reduced small--scale power. For
example, a single GRB at $z\gsim 10$, or 10 GRBs at $z\gsim 5$,
discovered by {\it Swift} during its scheduled two--year mission, would
rule out an exponential suppression of the power spectrum on scales
below $R_c=0.09$ Mpc (exemplified by warm dark matter models with a
particle mass of $m_x=2$ keV).  Models with a less sharp suppression
of small--scale power, such as those with a red tilt or a running
scalar index, $n_s$, are more difficult to constrain, because
they are more degenerate with an increase in the power spectrum
normalization, $\sigma_8$, and with models in which star--formation is
allowed in low--mass minihalos.  We find that a tilt of $\delta
n_s\approx 0.1$ is difficult to detect; however, an observed rate of 1
GRB/yr at $z \gsim 12$ would yield an upper limit on the running of the
spectral index, $\alpha\equiv dn_s/d\ln k > -0.05$.
\end{abstract}
\vspace{+0.5cm}
]

\section{Introduction}
\label{sec:intro}

In the years leading up to the recent launch of the {\it Swift}
satellite,\footnote{See http://swift.gsfc.nasa.gov} it has been
increasingly recognized that distant gamma ray bursts (GRBs) offer a
unique probe of the high redshift universe.  In particular, GRBs are
the brightest known electromagnetic phenomena in the universe, and can
be detected up to very high redshifts (e.g. Wijers et al. 1998; Lamb
\& Reichart 2000; Ciardi \& Loeb 2000), well beyond the redshift
$z\approx 6.5$ of the most distant currently known quasars (Fan et
al. 2003) and galaxies (Kodaira et al. 2003).

There is increasing evidence that GRBs are associated with the
collapse of short--lived, massive stars, including the association of
bursts with star--forming regions (e.g. Fruchter et al. 1999), a
contribution of supernova light to the optical afterglow (e.g. Bloom
et al. 1999; Garnavich et al. 2003), and most directly, association
with a supernova (Stanek et al. 2003; Hj\"orth et al. 2003).

As a result, the redshift distribution of bursts should follow the
mean cosmic star--formation rate (SFR).  Several studies have computed
the evolution of the expected GRB rate under this assumption, based on
empirical models of the global SFR (Totani 1997, 1999; 1999; Wijers et
al. 1998; Lamb \& Reichart 2000; Ciardi \& Loeb 2000).  Recent
determinations of the cosmic SFR out to redshift $z\sim 5$
(e.g. Bunker et al. 2004; Gabasch et al. 2004; Giavalisco et al. 2004)
show that star--formation is already significant at the upper end of
the measured redshift range, with $\gsim 10\%$ of all stars forming
prior to $z=5$, which would result in a significant population of GRBs
at these redshifts.  Further associating star--formation with the
formation rate of non--linear dark matter halos, and using theoretical
models based on the Press \& Schechter (1974) formalism, Bromm \& Loeb
(2002) and Choudhury \& Srianand (2002) have extrapolated the SFR and
obtained the GRB rates expected at still higher redshifts.  These
studies have concluded that a significant fraction (exceeding several
percent) of GRBs detected at {\it Swift}'s sensitivity should
originate at redshifts as high as $z>10$.  The spectra of bright
optical/IR afterglows of such distant GRBs can then reveal absorption
features by neutral hydrogen in the intergalactic medium (IGM), and
can serve as an especially clean probe of the reionization history of
the universe (Miralda-Escud\'e 1998; Lamb \& Reichart 2000; Choudhury
\& Srianand 2002; Lamb \& Haiman 2003; Barkana \& Loeb 2004).
 
In this paper, we investigate a different method to utilize distant
GRBs, and to glean information about early structure formation. The
mere presence of a GRB at, say, $z>10$ will indicate that non--linear
dark matter (DM) structures already exist at this redshift: the stars
that give birth to the GRBs must form out of gas that collected inside
dense DM potential wells.  Structure formation in a cold dark matter
(CDM) dominated universe is ``bottom--up'', with low--mass halos
condensing first.  In the current concordance cosmology, with
densities in cold dark matter (CDM) and dark energy of $(\Omega_{\rm
M},\Omega_{\rm \Lambda})\approx (0.3,0.7)$ that have emerged from {\it
WMAP} and other recent experiments (Spergel et al. 2003), DM halos
with the masses of globular clusters, $10^{5-6}\msun$ condense from
$\sim 3\sigma$ peaks of the initial primordial density field as early
as $z\sim 25$.  It is natural to identify these condensations as the
sites where the first astrophysical objects, including the first
massive stars, were born.  As a result, one expects to possibly find
GRBs out to this limiting redshift, but not beyond.

With a scale--invariant initial fluctuation power spectrum, the CDM
model has been remarkably successful, and has matched many observed
properties of large--scale structures in the universe, and of the
cosmic microwave background (CMB) radiation.  However, the power
spectrum on smaller scales, corresponding to DM halo masses of $M\lsim
10^9~{\rm M_\odot}$, remains poorly tested.  Recent observations
suggest that the standard model predicts too much power on small
scales: it predicts steep cusps at the centers of dark matter halos,
whereas the rotation curves of dwarf galaxies suggest a flat core; it
also predicts more small satellites than appear to be present in the
Local Group (these and other problems with CDM on small scales are
reviewed by, e.g. Sellwood \& Kosowsky 2001 and Haiman, Barkana \&
Ostriker 2001).  Although astrophysical explanations of these
observations are possible, much recent attention has been focused on
solutions involving the properties of dark matter.  Proposals include
self-interacting dark matter (Spergel \& Steinhardt 2000), a
repulsive interaction to gravity (Goodman 2000; Peebles 2000), the
quantum--mechanical wave properties of ultra--light dark matter
particles (Hu et al. 2000), and a resurrection of warm dark matter
(WDM) models (Bode et al. 2001).

By design, a common feature of models that attempt to solve the
apparent small--scale problems of CDM is the reduction of fluctuation
power on small scales.  In addition, we note that reduced small scale
power is a direct consequence of a range of slow--roll inflationary
models, which predict a red tilt of the power spectrum, $n_s\equiv
d\ln P(k) / d\ln k < 1$, and a running of the spectral index $\alpha\equiv
dn_s/d\ln k <0$ (see, e.g., Kinney 2003 for a general discussion of
power spectra predicted in different inflation models, and Kosowsky \&
Turner 1995 for a discussion of models with a running index).
Interest in such models was recently re--kindled, as they appeared
preferred by a combination of CMB anisotropy data from {\it WMAP} with
fluctuation statistics in the \lya forest (Spergel et al. 2003; Peiris
et al. 2003).

The loss of small--scale power generically suppresses structure
formation most severely at the highest redshifts, where the number of
self--gravitating objects is drastically reduced.  In each model,
there exists a redshift beyond which the number of GRBs (or any other
object) is exponentially suppressed, and a detection of a GRB beyond
this redshift can be used to constrain such models.
A similar constraint can be obtained from the observed reionization of
the universe at high--redshifts.  For example, Barkana, Haiman \&
Ostriker (2001; hereafter BHO) showed that in the case of WDM models,
invoking a WDM particle mass of $m_x\sim 1$ keV (approximately the mass
required to solve the problems listed above), the paucity of ionizing
sources makes it difficult to account for the reionization of the
universe by redshift $z\sim 6$. Reionization as early as $z\sim 17$,
as recently suggested by {\it WMAP} observations of CMB anisotropies
(Spergel et al. 2003) would require extreme efficiencies for star
formation and ionizing photon production for masses of $m_x\lsim 2$ keV
(see also Somerville, Bullock \& Livio 2003, who reach similar
conclusions).

GRBs, if discovered at $z>6$, have the potential to provide
independent and stronger constraints. {\it The purpose of this paper
is to quantify the constraints that the detection of distant GRBs
would place on structure formation models with reduced small--scale
power.}  Throughout most of our calculations, we focus on WDM model as
a proxy, but our results are valid for any theory which imposes a
small--scale cut--off in the primordial power spectrum.

The rest of the paper is organized as follows. In
\S~\ref{sec:mass_functions}, we briefly describe our Monte Carlo
approach to modify the standard Press-Schechter formalism, allowing us
to compute halo mass functions in WDM models.  In
\S~\ref{sec:GRB_rates}, we describe our method to compute the GRB
rates using the halo mass functions, and simple models for the
intrinsic GRB luminosity function.  In \S~\ref{sec:constraints}, we
present the constraints that high redshift GRB detections would place
on the small--scale power spectrum.  Finally, in
\S~\ref{sec:conclusions}, we discuss the implications of this work and
offer our conclusions.

Unless stated otherwise, throughout this paper we assume standard
cosmological parameters, ($\Omega_\Lambda$, $\Omega_{\rm M}$,
$\Omega_b$, n, $\sigma_8$, $H_0$) = (0.73, 0.27, 0.044, 1, 0.85, 71 km
s$^{-1}$ Mpc$^{-1}$), consistent with {\it WMAP} measurements of the
CMB power spectrum on large scales (Spergel et al. 2003), and quote
all lengths in comoving units.

\section{Mass Functions in CDM and in WDM}
\label{sec:mass_functions}

In this section, we briefly review the DM halo mass functions obtained
in the Press-Schechter (1974; hereafter PS) and extended
Press--Schechter (see Lacey \& Cole 1993; hereafter EPS) formalisms,
together with
the modifications required to model structure formation in CDM models
with reduced small--scale power. Our treatment closely follows that of
BHO (which the reader is encouraged to consult for more details).  In
\S~\ref{sec:eps}, we describe the standard mass function, and in
\S~\ref{sec:cutoff}, we motivate the parameterization of a power
spectrum cut--off in the WDM model. In \S~\ref{sec:mc}, we describe our
Monte Carlo simulations needed to incorporate the additional effective
pressure of the WDM particles with non--negligible velocity
dispersion.  Readers not interested in the modeling details of the DM
halos can skip directly to the \S~\ref{sec:GRB_rates}, which describes
how we associate GRBs with these halos.

\subsection{Halo Formation in CDM}
\label{sec:eps}

The mass function of DM halos in CDM models can be derived in closed
form in the PS formalism.  Although the PS mass function is in fair
agreement with simulations, especially for the ``typical'' halos, it
underestimates the number of rare, massive halos that are most
relevant for our purposes; it also overestimates the number of
low--mass halos, when compared with large numerical simulations
(e.g. Jenkins et al. 2001).  Here we adopt a modified expression
obtained by Sheth \& Tormen (1999), which fits the simulated mass
function to an accuracy of $\sim 10\%$,
\begin{equation}
\label{eq:st_dndM}
\frac{dn(>M,z)}{dM} = - \frac{\langle \rho \rangle}{M} \frac{\partial (\ln ~ \sig)}{\partial M} \sqrt{\frac{2}{\pi}}
A \left(1 + \frac{1}{\hat{\nu}^{2p}} \right) \hat{\nu} \exp \left[ - \frac{\hat{\nu}^2}{2} \right].
\end{equation}
Here, $dn/dM$ is the comoving number density of halos per unit mass, $M$ is the total mass of the halo, $\langle \rho
\rangle=\Omega_{\rm M} \rho_{\rm crit}$ is the mean background
matter density,
\begin{equation}
\label{eq:ps_sig}
\sigsq = \int_{0}^{\infty} \frac{k^2 dk}{2 \pi^2} P(k)^2 W_k^2(M),
\end{equation}
is the {\it r.m.s.} fluctuation in the mass enclosed within a region
described by a top--hat filter in real space $W(M)$ (and its Fourier
transform $W_k$), and $\hat{\nu} \equiv \sqrt{a} \delta_c(z) / \sig$,
where $a$, $p$, and $A$ are fitting parameters.  Sheth, Mo \& Tormen
(2001) derive this form of the mass function by including shear and
ellipticity in modeling non--linear collapse, effectively changing the
scale-free critical over--density $\delta_c(z)\approx 1.68$, obtained
in the case of spherically symmetric collapse (Peebles 1980), into a
function of filter scale,
\begin{equation}
\label{eq:st_delc}
\delc = \sqrt{a} \delta_c(z) \left[ ~ 1 + b \left(\frac{\sigsq}{a \delta_c^2(z)}\right)^c ~ \right].
\end{equation}
Here $b$ and $c$ are additional fitting parameters ($a$ is the same as
in eq.~\ref{eq:st_dndM}).  We use this correction to obtain the
critical threshold $\delc$ from $\delta_c(z)$ (in the WDM case,
$\delta_c(z)$ itself is modified as described in \S~\ref{sec:mc}
below).  For the constants in equations (\ref{eq:st_dndM}) and
(\ref{eq:st_delc}), we adopt the recent values obtained by Jenkins et
al. (2001), who studied a large range in redshift and mass: $a$ =
0.73, $A$ = 0.353, $p$ = 0.175, $b$ = 0.34, $c$ = 0.81.

\subsection{Power Spectrum Cut--Offs}
\label{sec:cutoff}
 
Structure formation in WDM matter models differs from CDM in two main
ways: (1) the free-steaming velocities of the particles wash--out
small--scale overdensities; (2) the residual particle velocities,
although they redshift away as $(1+z)$, create an effective
``pressure'' which slows the early growth of perturbations.  Both
effects suppress structure formation on small scales, but for the sake
of generality we discuss the two effects separately.  Free streaming
is easily included computationally as a suppression of the power
spectrum of fluctuations, and it is qualitatively similar, for
example, to changes in the inflationary potential, which determines
the power spectrum.  Likewise, a red tilt of the power spectrum,
$n_s\equiv d\ln P(k) / d\ln k < 1$, or a running of the spectral index,
$\alpha\equiv dn_s/d\ln k <0$, is easily included in the analysis by
simply modifying the power spectrum in equation~\ref{eq:ps_sig}.  On
the other hand, the effective pressure in WDM models is more difficult
to include computationally, and is specific to the WDM model; it is
discussed separately in the next subsection.

Free streaming manifests itself as a cut--off in the power-spectrum,
which ``flattens'' $\sigma(M)$ for small $M$.  This effect becomes more
severe as the WDM particle mass is lowered, as demonstrated in the
bottom panel of Figure~\ref{fig:delc_sig} for $m_x=2$ and 1 keV.  To
produce a given energy density contribution, $\Omega_x$ (where we take
$\Omega_x$ $\equiv$ $\Omega_{\rm M} - \Omega_b$), the required WDM
particle mass is determined by $m_x n_x$ $\propto$ $\Omega_x h^2$,
where the present number density, $n_x$, follows from the particle
distribution function.  This can be used (see BHO) to relate the
particle mass and the r.m.s. velocity dispersion, $v_{rms}$:
\begin{equation}
v_{rms}(z) = 0.0437 (1 + z) \left( \frac{\Omega_x h^2}{0.15}
\right)^{1/3} \left( \frac{g_x}{1.5} \right)^{-1/3} \left(
\frac{m_x}{\rm 1 ~ keV} \right)^{-4/3},
\end{equation}
where $g_x$ is the effective number of degrees of freedom of WDM.  The
usual assumption of a fermionic spin-$\frac{1}{2}$ particle yields
$g_x$ = 1.5.  This modifies the CDM power spectrum (which we take from
Eisenstein \& Hu 1999) by multiplying it with the square of a transfer
function (Bode, Ostriker, \& Turok 2001):
\begin{equation}
\label{eq:tf}
T_x(k) = ( 1 + (\epsilon k R_c)^{2 \nu} )^{- \eta / \nu},
\end{equation}
with parameters $\epsilon$ = 0.361, $\eta$ = 5, $\nu$ = 1.2.  The
power spectrum is reduced to half its value in CDM at $k = 1/R_c$,
where the cut--off scale, $R_c$, is given by
\begin{equation}
\label{eq:R_c}
R_c = 0.201 \left( \frac{\Omega_x h^2}{0.15} \right)^{0.15} \left(
\frac{g_x}{1.5}\right)^{-0.29} \left( \frac{m_x}{\rm 1 ~ keV}
\right)^{-1.15},
\end{equation}
where $R_c$ is in comoving Mpc.  Here, we only consider particles with
fermionic spins, i.e. $g_x$ = 1.5, but all results can be scaled for
arbitrary values of $g_x$ using the equation above. The corresponding
mass scale, $M_c = (4/3) \pi R_c^3 \langle \rho \rangle$, is
\begin{equation}
M_c = 1.74 ~ \times ~ 10^8 ~ \left( \frac{\Omega_{\rm M} h^2}{0.15} \right) \left( \frac{R_c}{0.1 ~ {\rm Mpc}} \right)^3 ~ {\rm M_\odot} ~ .
\end{equation}
We will use $R_c$ and $m_x$ interchangeably, with equation
(\ref{eq:R_c}) relating them.

\vspace{+0\baselineskip} \myputfigure{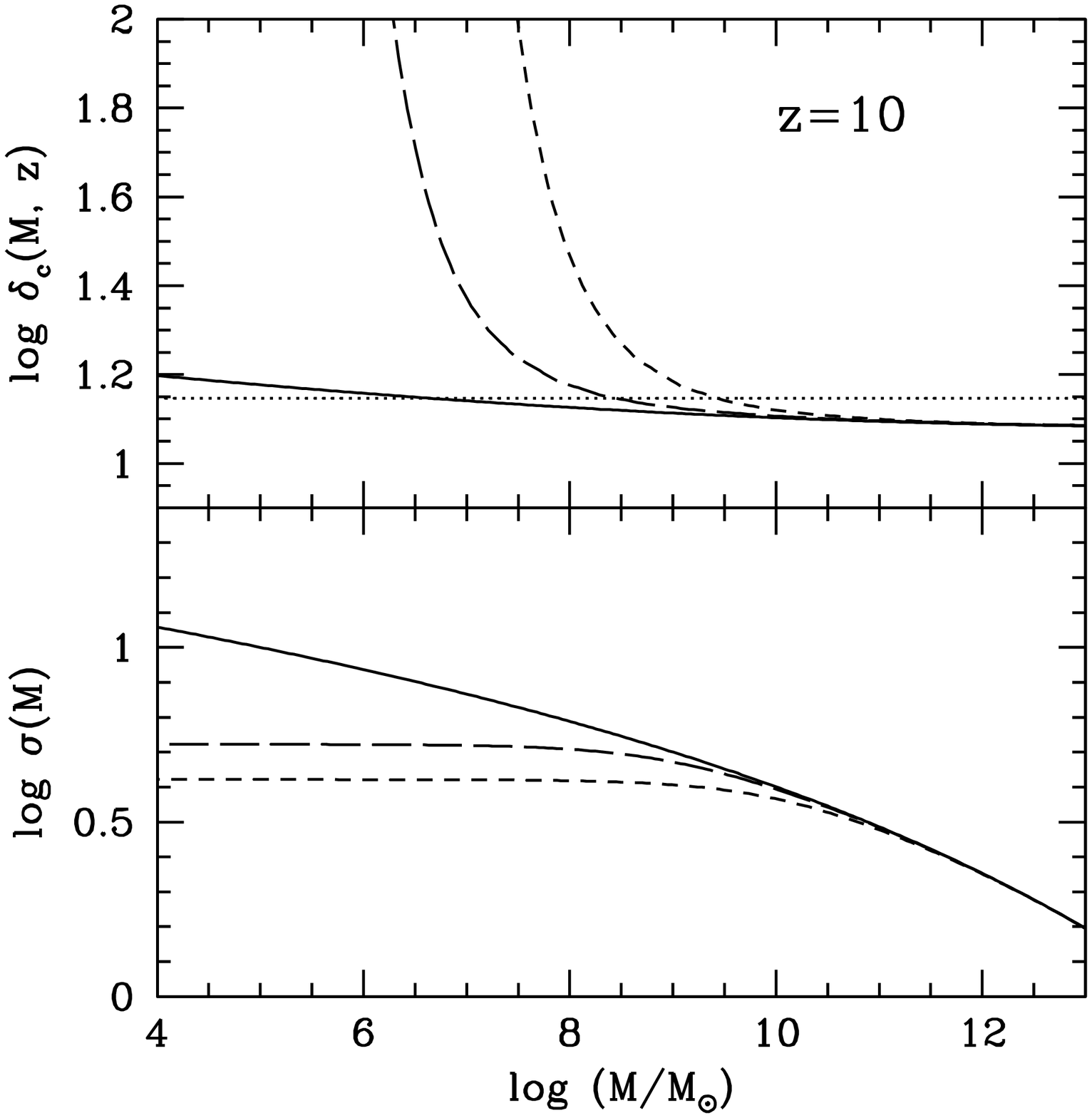}{3.3}{0.5}{.}{0.}
\vspace{-1\baselineskip} \figcaption{ {\it Top panel}: critical
overdensity threshold as a function of scale, evaluated for $z=10$.
The {\it dotted curve} represents the scale--free critical overdensity
$\delta_c(z)$.  The {\it solid curve} corresponds to CDM, the {\it
long--dashed curve} correspond to a WDM particle mass of $m_x$ = 2 keV
, and the {\it short--dashed curves} correspond to $m_x$ = 1 keV .
All curves incorporate the correction from
equation~(\ref{eq:st_delc}), except for the dotted curve.  {\it Bottom
panel}: the {\it r.m.s.}  mass fluctuation in top--hat filters of
mass--scale $M$.  As above, the {\it solid curve} corresponds to CDM,
the {\it long--dashed curve} corresponds to $m_x$ = 2 keV (or
power--spectrum cut--off scale $R_c$ = 0.087 Mpc), and the {\it
short--dashed curve} corresponds to $m_x$ = 1 keV (or $R_c$ = 0.193
Mpc).
\label{fig:delc_sig}}
\vspace{+1\baselineskip}

\subsection{Effective Pressure of WDM Particles}
\label{sec:mc}

As mentioned above, structure formation in WDM models is further
suppressed by the residual velocity dispersion of the WDM particles,
which delay the growth of perturbations.  We use the results of BHO,
who made an analogy with an ideal gas whose temperature corresponds to
the velocity dispersion of the WDM.  In this case, the pressure delays
the collapse, and can be effectively included in the mass function
computed in the EPS analysis, by raising the critical linear
extrapolated overdensity threshold at collapse $\delta_c(z)$.  BHO
computed the critical overdensity by following the collapse of
spherical perturbations, using a one-dimensional, spherically
symmetric Lagrangian hydrodynamics code, originally developed by Thoul
\& Weinberg (1995).  The top panel in Figure~\ref{fig:delc_sig} shows
$\delta_c$ (adopted from BHO) as a function of $M$, at the fixed
redshift $z=10$.  The dotted curve represents the scale--free critical
overdensity $\delta_c(z)$ arising from spherical collapse. The solid
curve corresponds to CDM, the long--dashed curve corresponds to WDM
with $m_x$ = 2 keV , and the short--dashed curve corresponds to $m_x$
= 1 keV.  The threshold is found to increase sharply below the mass
scale
\begin{eqnarray}
\nonumber M_J = &3.06&~\times~ 10^8~\left(\frac{g_x}{1.5}\right)^{-1} \left(\frac{\Omega_x h^2}{0.15}\right)^{1/2}\\
&\times& ~ \left(\frac{m_x}{\rm 1~keV}\right)^{-4} \left(\frac{1+z_i}{3000}\right)^{3/2} ~ {\rm M_\odot},
\end{eqnarray}
which can be shown (see BHO) to correspond to an analog of a ``Jeans
mass'', i.e. the scale of the objects whose collapse is significantly
delayed by the pressure.

\vspace{+0\baselineskip}
\myputfigure{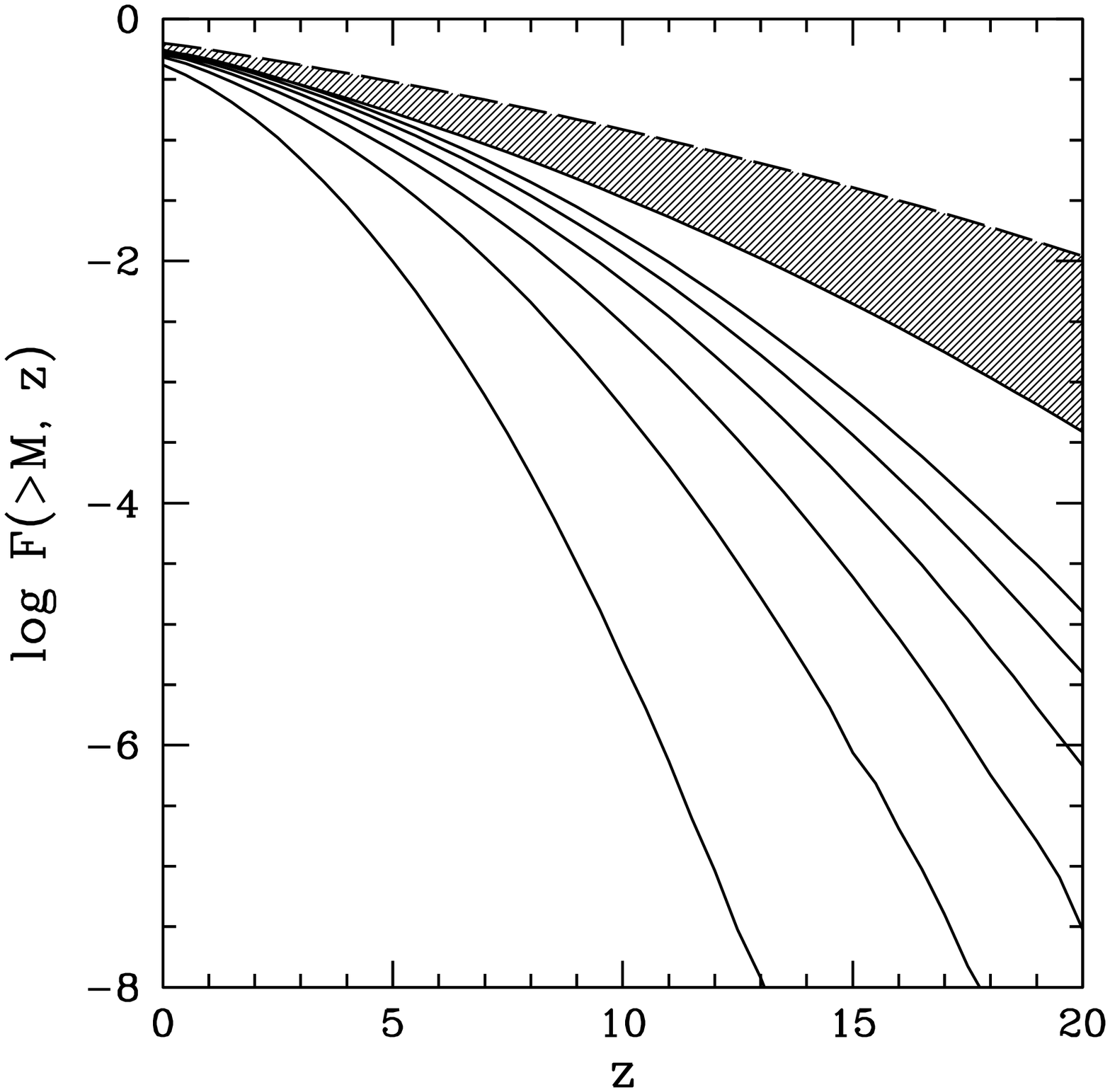}{3.3}{0.5}{.}{0.}  
\vspace{-1\baselineskip} \figcaption{ The fraction of the total mass
collapsed into halos of mass $M$ or higher, $F(>M,z)$, as a function
of redshift.  The shaded region shows the collapsed fraction in CDM,
with a range of low--mass cutoffs corresponding to virial temperatures
300 K $< T_{\rm vir} <$ $10^4$ K.  The other curves correspond to WDM
particle masses of $m_x$ = 3.0 keV, 2.5 keV, 2.0 keV, 1.5 keV, 1.0
keV, 0.5 keV, from top to bottom, and do not assume any low--mass
threshold (effectively, $M=0$).  The cutoffs used in the CDM case
would leave the WDM results essentially unchanged (see discussion in
text).  \label{fig:mfvsz}}
\vspace{+1\baselineskip}

Since $\delc$ is function of scale in WDM models, one can not obtain
WDM mass functions from the standard EPS analysis, which makes use of
the symmetry in the random walk trajectories of $\delta$ vs. $M$ about
a fixed threshold (Lacey \& Cole 1993).  Instead, we compute the mass
functions using Monte--Carlo simulations.  We generate random
realizations of trajectories $\del$, as the scale is decreased from $
M\sim \infty$, and generate the histograms of the scales at which the
trajectory first crosses the $\delc$ threshold.  Each step in the
random walk, $\Delta\del$, is Gaussian distributed with a variance of
$\sigma_{\rm step}^2(M, \Delta M)$ = $\sigsq$ - $\sigma^2(M - \Delta
M)$.  When constructing such a random walk, one must be careful to use
steps small enough such that as the smoothing scale $M$ is decreased
by $\Delta M$, the likelihood that the $\delc$ threshold is crossed
between $M$ and $M-\Delta M$ is small.  The physical reason for this
is the so--called ``cloud-in-cloud problem'': to ensure that we do not
step over a collapsed halo as we decrease the smoothing scale $M$
(i.e. that our $\del$ trajectory has not gone above $\delc$ and then
dropped below it again within the range $\Delta M$).  We use an
adaptive step--size, set so that the barrier $\delc$ is at least $7
\sigma_{\rm step}$ away from $\del$, with a minimum resolution of
$\Delta M$ = $M/100$.  Formally, defining $\Delta_7M$ such that
$(\delc - \del)$ / $\sigma_{\rm step}(M, \Delta_7M)=7$, our step size
is $\Delta M$ = MAX[$M$/100, $\Delta_7M$].  We find that these
parameters efficiently reproduce the standard EPS mass function in the
CDM case to an accuracy of a few percent.

We also find that a starting mass for the random walk trajectories as
small as $\sim 10^{16}~{\rm M_\odot}$ is sufficient to obtain accurate
mass functions at $z = 0$.  The starting mass can be decreased as
redshift increases, since the characteristic mass that is collapsing
gets smaller as redshift increases.  A starting mass of $\sim
10^{12}~{\rm M_\odot}$ is sufficient at $z = 15$.

The number of simulated trajectories required to obtain accurate mass
functions is a strong function of redshift.  This is to be expected,
since virialized structure becomes very rare at high redshift.  We
show this effect in Figure \ref{fig:mfvsz}, where we plot the fraction
of the total mass collapsed into halo of mass $M$ or higher,
$F(>M,z)$, as a function of redshift.  The shaded region shows the
collapsed fraction in CDM, with a range of low--mass cutoffs
corresponding to virial temperatures 300 K $ < T_{\rm vir} < 10^4$ K (see
discussion in \S~\ref{sec:GRB_rates} below).  The other curves
correspond to WDM particle masses of $m_x$ = 3.0 keV, 2.5 keV, 2.0
keV, 1.5 keV, 1.0 keV, 0.5 keV, from top to bottom, and do not assume
any low--mass threshold (effectively, $M=0$).  Introducing the same
two low--mass cutoffs as in the CDM case would leave the WDM results
essentially unchanged, since the power is already strongly suppressed
in excess of these cutoffs at the high redshifts where these low mass halos
would dominate the collapse fraction (see discussion below, or Fig. 7
in BHO).  As can be seen, simulating high redshift mass functions for
small particle masses can be prohibitively computationally expensive.
For example, we find that accurate mass functions for $m_x < 1$ keV at
$z > 15$, require $\gsim 10^9$ Monte--Carlo runs, as less than one in a
million $\del$ trajectories crosses $\delc$.

\vspace{+0\baselineskip} \myputfigure{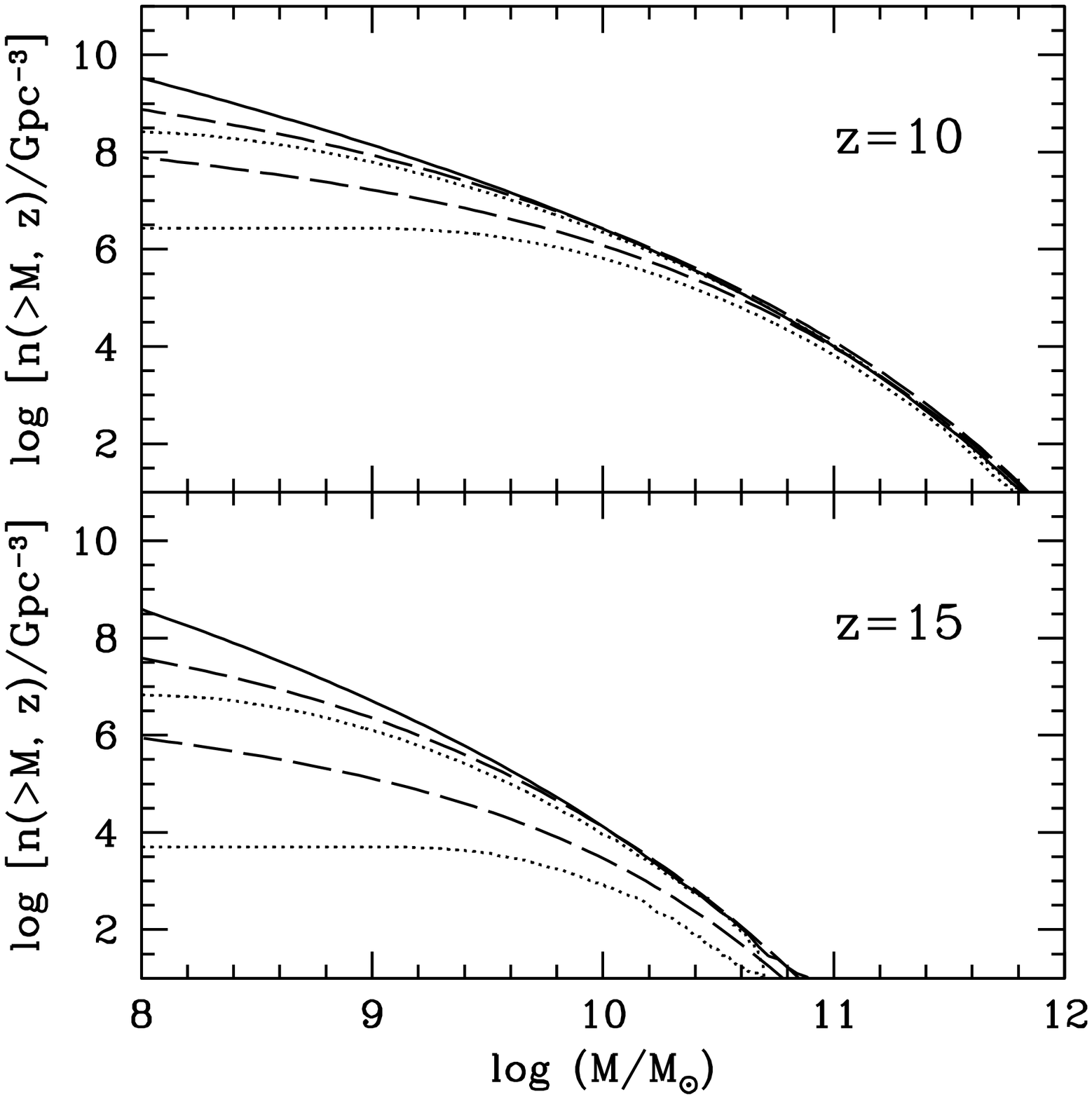}{3.3}{0.5}{.}{0.}
\vspace{-1\baselineskip} \figcaption{Cumulative mass functions (number
of halos with masses greater than $M$ per $\rm Gpc^{3}$) at redshifts
$z$=10 ({\it top panel}) and $z$=15 ({\it bottom panel}).  The {\it
solid curves} show mass functions in CDM models; the {\it dotted
curves} correspond to WDM models with $m_x$ = 2 keV and 1 keV (from
top to bottom); the {\it dashed curves} are mass functions in the same
two WDM models, but incorporating only a power spectrum cut--off
(\S~\ref{sec:cutoff}), and no WDM pressure (\S~\ref{sec:mc}).  The WDM
cumulative mass functions were created with $10^9$ Monte--Carlo
realizations of $\del$ trajectories.
\label{fig:mf_z10_z15}}
\vspace{+1\baselineskip}

Figure~\ref{fig:mf_z10_z15} shows sample cumulative mass functions at
redshifts $z$=10 ({\it top panel}) and $z$=15 ({\it bottom panel}).
The solid curves correspond to CDM; the dotted curves correspond to
WDM models with $m_x$ = 2 keV and 1 keV (from top to bottom).  The
dashed curves are mass functions for the same two WDM models, but
incorporate only a power spectrum cut--off, ignoring the effective
pressure of WDM. The WDM mass functions were created with $10^{9}$
Monte-Carlo runs as explained above.  (Note that our results are
slightly different from those of BHO; this is due to a small correction
to BHO's derivation of $R_c$ and the corresponding $\sigma(M)$.)

As can be seen from Figure~\ref{fig:mf_z10_z15}, including the
pressure term in the WDM models further suppresses the number of halos
relative to the models which include only the power spectrum cut--off.
Furthermore, the relative importance of the pressure term increases
with increasing redshift. As anticipated, the overall differences
between the WDM and CDM mass functions increase toward higher
redshift.  This is because in the early universe, the characteristic
scale of collapsing and virializing halos was smaller, and closer to
the cut--off scales discussed above.  These large differences will aid
in discriminating between models with different power spectrum cut--off
scales.

\section{The Evolution of the GRB Rate with Redshift}
\label{sec:GRB_rates}

In this section, we describe a model for the expected evolution in the
rate of all GRBs, as well as the fraction that can be detected by {\it
Swift}.  As is commonly done in the context of predicting the rate of
GRBs at high redshift (Bromm \& Loeb 2002; Choudhury \& Srianand
2002), we assume that the GRB rate density (the number of GRBs per
unit time per unit comoving volume), $\GRBrd$, is proportional to the
global star--formation rate density,
\begin{equation}
\label{eq:GRBrd_k_sfrd}
\GRBrd ~ \approx ~ K ~ \sfrd ~ ,
\end{equation} 
where $K$ is the proportionality constant in units of $M_\odot^{-1}$,
and $\sfrd$ is the stellar mass produced on average per unit comoving
volume per unit time.  First we discuss the evolution of the star
formation rate, and then the normalization of the corresponding GRB
rate.  In \S~\ref{sec:exp} we discuss the uncertainties of our approach.

\subsection{Star--Formation Rate in Halos}
\label{sec:sfr}

We estimate the global star--formation rate (SFR) density at redshift
$z$ as
\begin{equation}
\label{eq:sfrd_general}
\sfrd = \epsilon_\ast \frac{\Omega_b}{\Omega_{\rm M}} \int_{M_{\rm min}}^{\infty} dM \int_{\infty}^{z} dz' M \frac{\partial^2 n(>M, z')}{\partial M \partial z'} P(\tau) ~ ,
\end{equation}
where $\epsilon_\ast$ is the efficiency parameter for the conversion
of gas into stars, $M$ $dM$ $\partial n(>M, z') / \partial M$ is the mass density contributed by halos with masses between $M$ and $M + dM$ at redshift $z$, $t(z)$ is the age of the universe at redshift $z$, and $P(\tau)$ is the probability per unit time that new stars form in a mass element of age $\tau \equiv t(z) - t(z')$.  We adopt the fiducial value of $\epsilon_\ast = 0.1$ (see, e.g., Cen [2003]), but note that our results are insensitive to this value since we normalize the coefficient $K$ in order to match our total GRB rate with observations
(the important assumption is only that $\epsilon_\ast$ is
constant; this assumption is probably conservative, as discussed
below). We also note that the simple
time--derivative of the halo mass--function in
equation (\ref{eq:sfrd_general}), in general, contains a contribution
from mergers between halos, in addition to the formation of new halos;
see, e.g., Sasaki 1994 for a discussion.  However, the integral in
equation (\ref{eq:sfrd_general}) is sensitive only to the total
virialized mass above mass $M_{\rm min}$, so that this ambiguity
should not affect our results (although we note that the formation of
new halos dominates in our case since the relevant halos are above the
non--linear mass--scale).

The minimum mass, $M_{\rm min}$, depends on the efficiency of gas
cooling and collapsing into a dark matter halo.  Without molecular
hydrogen, $M_{\rm min}$ corresponds to a halo with virial temperature,
$T_{\rm vir} \sim 10^4$ K; with $\rm H_2$, $T_{\rm vir} \sim 300$ K
(Haiman, Abel, \& Rees 2000; we use the conversion between halo mass
and virial temperature as given in Barkana \& Loeb 2001).  The amount
of $\rm H_2$ present in the early universe is uncertain, so below we
present results for both $T_{\rm vir} > 10^4$ K and $T_{\rm vir}
> 300$ K.  We note that only the CDM mass functions are sensitive
to these cooling thresholds (see Fig. \ref{fig:sfrd_fract}).  In our
models which incorporate small--scale power suppression, as the
characteristic collapse scale approaches the cooling cut--off, $T_{\rm
vir}$, the cooling cut--off has already become smaller than the power
spectrum cut--off, $R_c$.  Hence, we do not distinguish between models
with $T_{\rm vir} > 10^4$ K and $T_{\rm vir} > 300$ K for the
WDM particle masses presented below.

In the presence of an ionizing background, the cosmological Jeans
mass, above which gas can collect in DM halos, is increased,
corresponding to virial temperatures $\gsim 10^4$ K (Rees 1986;
Efstathiou 1992).  Early work on this subject (Thoul \& Weinberg 1996)
suggested that an ionizing background would completely suppress star
formation in halos with circular velocities $v_{\rm circ} \lsim 35$
km/s, and partially suppress star--formation in halos with 35 km/s
$\lsim v_{\rm circ} \lsim$ 93 km/s.  More recently, Dijkstra et
al. (2004) found that such a suppression is likely to be countered by
strong self--shielding for $z \gsim$ 3.  For concreteness, we
completely suppress star--formation in halos with $T_{\rm vir} < 10^4$
K for $z < 7$ (assumed to correspond to the redshift of reionization,
when the background radiation is established; e.g. Mesinger \& Haiman
2004), and increase the cut--off to $v_{\rm circ} < 55$ km/s ($T_{\rm
vir} < 1.1\times 10^5$ K) for $z < 3$ (when the intermediate halos
with 10 km/s $< v_{\rm circ} < 55$ km/s are assumed to cease
self--shielding).  As can be seen in Figure~\ref{fig:sfrd_fract}
below, these assumptions only weakly affect our results, because most
of the contribution to $\sfrd$ at these redshifts comes from more
massive halos.  Furthermore, we already bracket a large range of
expected GRB distributions by presenting results for both $T_{\rm vir}
> 10^4$ K and $T_{\rm vir} > 300$ K.

Finally, we describe two different functional forms for the stellar
formation probability density $P(\tau)$.

\subsubsection{Instantaneous Star--Formation}
\label{sec:delta}

The simplest assumption is to adopt $P(\tau)$ to be a Dirac delta function,
\begin{equation}
P(\tau)  = \delta(\tau).
\end{equation}
This essentially assumes that a fraction $\epsilon_\ast
(\Omega_b/\Omega_{\rm M})$ of the currently virializing mass is
instantaneously converted into stars, and that previously virialized
mass does not contribute to $\sfrd$.  With this assumption, equation
(\ref{eq:sfrd_general}) becomes
\begin{equation}
\label{eq:sfrd_delta}
\sfrd = \epsilon_\ast \frac{\Omega_b}{\Omega_{\rm M}} \int_{M_{\rm min}}^{\infty} M \frac{dz}{dt} \frac{\partial^2 n(>M, z)}{\partial z \partial M} dM ~ .
\end{equation}
Since the WDM mass functions are obtained with computationally
expensive Monte-Carlo simulations, we only present their results for
this simple model; results for mass functions incorporating a power
spectrum cut--off are presented for both $P(\tau)$ models.  In
numerically calculating derivatives of the mass functions in WDM
($\Delta n(>M,z) / \Delta z$), we use $z$ step--sizes of $\Delta z
\sim$ 0.2 -- 0.5 (at high redshifts $\Delta n(>M,z) / \Delta z$ is a
flatter function of $z$, so good accuracy can be achieved even with a
larger $\Delta z$ step--size).  The resulting SFR densities are shown
in the top left panel of Figure~\ref{fig:sfrd_fract}.  The solid
curves assume $T_{\rm vir} > 10^4$ K, and include only the effect of
the power--spectrum cutoff, assuming cut--off scales of $R_c$ = 0 Mpc
($m_x \rightarrow \infty$; i.e. standard CDM), $R_c$ = 0.193 Mpc ($m_x
= 1$ keV), and $R_c$ = 0.087 Mpc ($m_x = 2$ keV), from top to bottom.
The dashed curves assume $T_{\rm vir} > 300$ K, with the same
cut--off scales.  The dotted lines include the additional effect of the
effective pressure of WDM particles with a mass of $m_x$ = 2 keV and
$m_x$ = 1 keV, from top to bottom.  Since power is strongly suppressed
on scales in excess of $T_{\rm vir} > 10^4$ K for these WDM
particle masses cut--off scales, results for $T_{\rm vir} > 10^4$ K
and $T_{\rm vir} > 300$ K are virtually the same.  In CDM models,
the range of virial temperatures result in a wider range of expected
SFR densities, highlighted by the shaded area.  Our results are within
the uncertainties of existing observational estimates of the SFR
density in the currently available, low-redshift ($z \lsim 2$)
regime (e.g. Schiminovich et al. 2004).  Our SFR density increases out
to $z\gsim 6$ (and to still higher redshift in the $T_{\rm vir} > 300$ K case), and is consistent with recent estimates at $2\lsim
z\lsim 6$ from the Hubble Ultra Deep Field (Bunker et al. 2004), the
GOODs surveys (Giavalisco et al. 2004) and the FORS Deep Field on the
VLT (Gabasch et al. 2004), after they incorporate a factor of 5--10
increase in the SFR (Adelberger \& Steidel 2000) due to dust
obscuration.  Note that the peak of the SFR density falls between
$3\lsim z\lsim 12$ in our range of models, and that the SFR remains
significant (exceeding its present day value) out to redshifts as high
as $z\gsim 25$.

We stress that instant star--formation is the most conservative model
in setting observational limits to a power spectrum cut--off, as it
predicts the largest number of high--redshift GRB detections.  Any
other, more realistic form for the stellar formation probability
density, $P(\tau)$, will delay GRB events, thereby smearing the event
rate toward lower redshifts, and decreasing the likelihood of
detecting high--redshift GRBs, especially in the presence of
small--scale power suppression (see \S~\ref{sec:abs_constraints}).

\subsubsection{Exponential Approximation for $P(\tau)$}
\label{sec:exp}
We alternatively assume that stellar formation occurs on a time-scale
corresponding to the dynamical time, $t_{dyn} \sim (G \rho)^{-1/2}$,
(Cen \& Ostriker 1992; Gnedin 1996):
\begin{equation}
P(t(z) - t(z')) = \frac{t(z) - t(z')}{t_{dyn}^2} \exp \left[ -
\frac{t(z) - t(z')}{t_{dyn}} \right] ~ ,
\end{equation}
where $\rho(z) \approx \Delta_c \rho_{\rm crit}(z)$ is the mean mass
density interior to collapsed spherical halos (e.g. Barkana \& Loeb
2001), and $\Delta_c$ is obtained from the fitting formula in Bryan \&
Norman (1998), with $\Delta_c$ = $18\pi^2$ $\approx$ 178 in the
Einstein--de Sitter model.

Since there is no unique,
physically-motivated, and self-consistent way to track individual mass
elements and halo mergers in the EPS formalism, assigning an age,
$\tau \equiv t(z) - t(z')$, to each mass element is somewhat
arbitrary.  The problem arises since two neighboring mass elements
which are part of the same halo, can in EPS be flagged as belonging to
two different halos with different masses (Somerville \& Kollatt 1999;
see also Sheth \& Pitman 1997 and Benson et al. 2004).
Here, we assume that a mass element ``starts its clock'' ($\tau=0$)
when it first becomes part of a halo with mass $M > M_{\rm min}$, and
that it carries around that clock through any future mergers without
resetting it.  This ambiguity is bypassed in \S~\ref{sec:delta}, since
$P(\tau)$ is assumed to be a delta function.

We present our SFR densities for this model in the top right panel of
Figure \ref{fig:sfrd_fract}.  The curves correspond to the same models
as shown in the top left panel, except the WDM models with pressure
(dotted curves) are not included.  As expected, introducing a finite
width to the stellar formation probability density preferentially
suppresses high--redshift star--formation.

We remark that the sharp drop at $z = 7$ in the $T_{\rm vir} > 300$ K,
CDM curve is an indicator of the assumed sudden reionization (Bromm \&
Loeb 2002; Choudhury \& Srianand 2002).  This feature is a direct
prediction of reionization models, and is indeed likely to be sharp
(see, for example, Cen \& McDonald 2002 for a discussion, and a claim
for detecting a related feature in the Lyman $\alpha$ transmission
spectrum of distant quasars). Given a sufficient number of GRBs, the
shape and location of such a feature can be mapped out, and used as a
probe of the epoch of reionization.  We postpone a detailed
exploration of the detectability of such a feature to a future paper.

\vspace{+1\baselineskip} \myputfigure{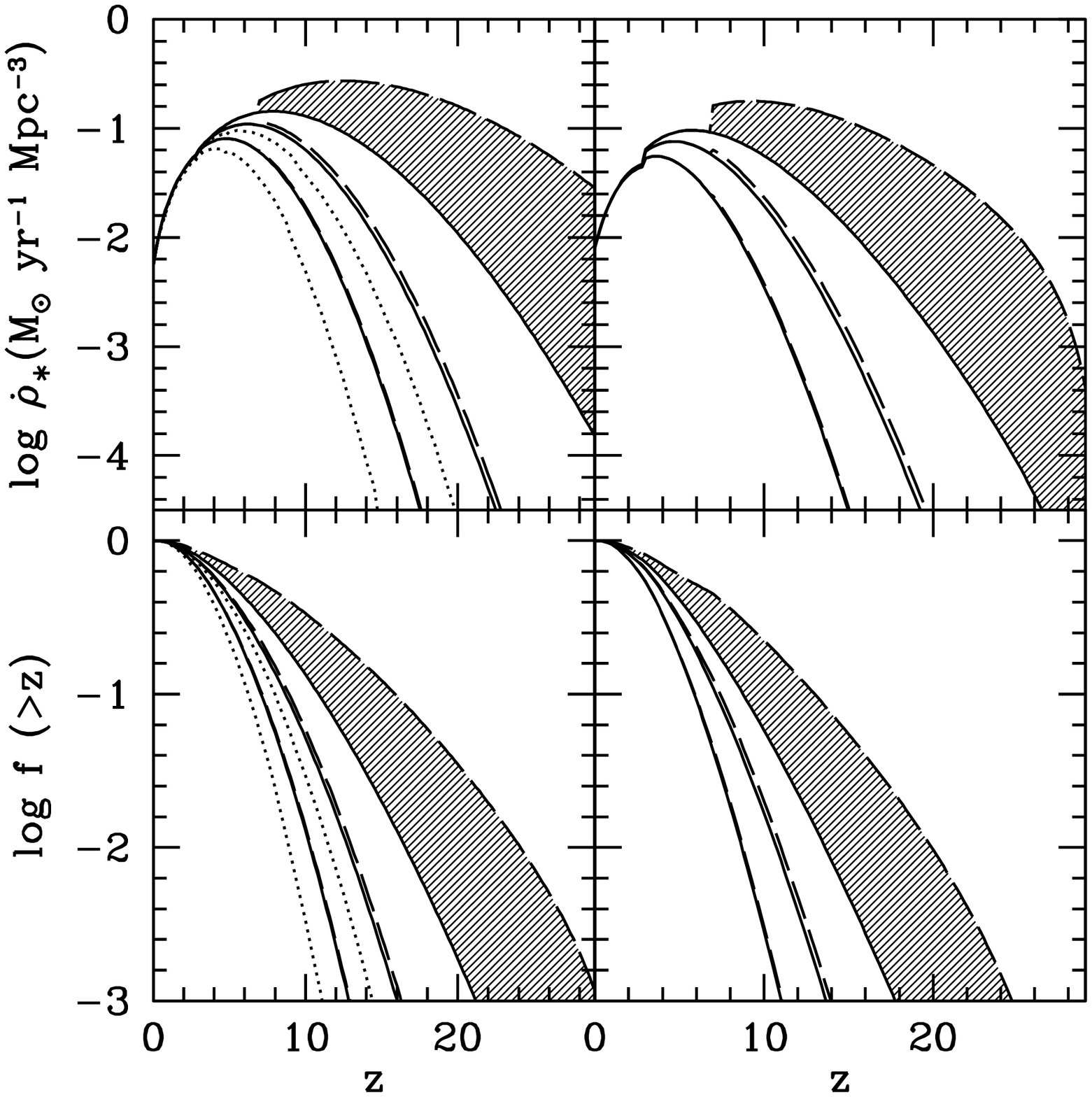}{3.3}{0.5}{.}{0.}
\vspace{-1\baselineskip} \figcaption{ {\it Top Panel:} SFR densities
assuming instantaneous star--formation in a collapsing halo ({\it left
panels}) and a finite exponential spread in star--formation times
({\it right panels}). The {\it solid curves} assume a minimum virial
temperature $T_{\rm vir}=10^4$ K of a halo for star--formation, and
include the effect of power--spectrum suppression below various
cut--off scales: $R_c$ = 0 Mpc ($m_x \rightarrow \infty$; i.e. standard
CDM), $R_c$ = 0.193 Mpc ($m_x = 1$ keV), and $R_c$ = 0.087 Mpc ($m_x =
2$ keV), from top to bottom.  The {\it dashed curves} assume a minimum temperature, $T_{\rm
vir}=300$ K, with the same cut--off scales.  The {\it dotted curves}
correspond to models that include the additional effect of the
effective pressure of WDM particles with mass $m_x$ = 2 keV and $m_x$
= 1 keV, from top to bottom.  The {\it shaded area} highlights the
expected range in SFR densities in CDM models with minimum virial
temperatures $300$K $\lsim T_{\rm vir}\lsim 10^4$K.  {\it Bottom
Panel:} The fraction of all GRBs that originate at redshifts higher
than $z$ in the models corresponding to the top panels.
\label{fig:sfrd_fract}}
\vspace{+1\baselineskip}

\subsection{The GRB Rate Associated with Star--Formation}
\label{sec:norm}

In order to constrain the proportionality constant $K$ by matching the
predicted and observed GRB rates, a luminosity function (LF) for the
GRB population needs to be assumed.  Observations have shown that GRBs
are far from being standard candles (and even if the total GRB energy
has a nearly universal value, the burst luminosity will vary; Frail et
al. 2001). However, the determination of their intrinsic LF has been
hampered so far by the lack of a sufficiently large sample with
detected redshifts.  On the other hand, fits to the observed flux
distribution suffer from a degeneracy between the LF and the SFR that
prevents an independent determination of both quantities.

On the theoretical side, simulations of jets in the collapsar model
(MacFadyen \& Woosley 1999) have shown that the jet energy (and hence
luminosity) is a decreasing function of the viewing angle $\theta$
relative to the jet axis, triggering studies of structured jets (Rossi
et al. 2002; Zhang \& M\'esz\'aros 2002). Here, following Rossi et
al. (2002), we assume that the luminosity has a form $L=L_{\rm
min}(\theta_{\rm jet}/\theta)^{2}$ for $\theta_{\rm core} < \theta <
\theta_{\rm jet}$, where $\theta_{\rm core}$ defines the core of the
jet (within which the luminosity is constant), and $\theta_{\rm jet}$
defines the outer edge of the jet (above which the luminosity drops to
zero).  Current observations have shown that $\theta_{\rm core}\lsim
0.06$ rad, and $\theta_{jet}\gsim 0.6$ rad (Bloom et al. 2003). Our
results are insensitive to the precise value of $\theta_{\rm core}$
below the observed minimum value, and therefore we set it to zero for
simplicity. The GRB rate at $z=0$, is, however dependent on
$\theta_{\rm jet}$ (or equivalently, the inferred minimum luminosity
$L_{\rm min}$; see also Guetta et al. 2004).

\vspace{+0\baselineskip} \myputfigure{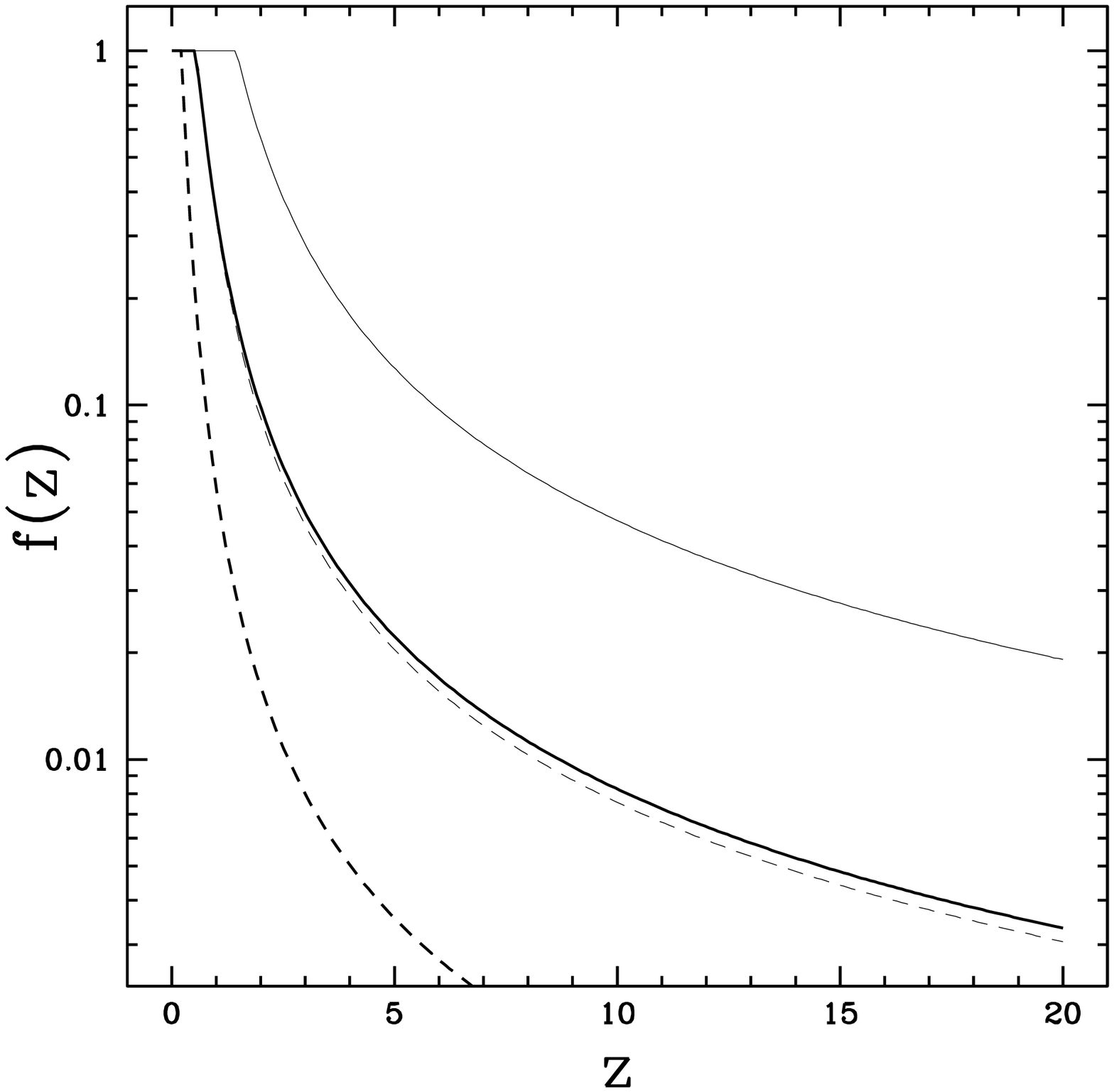}{3.3}{0.5}{.}{0.}
\vspace{-1\baselineskip} \figcaption{The fraction of all GRBs that are
brighter in our model luminosity function than the detection
thresholds of {\it BATSE} (dashed curves) and {\it Swift} (solid
curves).  The lower/upper curve in both cases corresponds to jet
angles of $\theta_{\rm jet}=\pi/2$ and 0.6 rad, respectively.  In each
case, we determine the minimum burst luminosity $L_{\rm min}$ so that
the flux distribution predicted in our CDM model with a threshold
virial temperature $ T_{\rm vir}=10^4$K best fits the distribution
observed by {\it BATSE}.
\label{fig:fractions}}
\vspace{+1\baselineskip}

For each of the models for the SFR described in the previous section,
we determine the constant $L_{\rm min}$ by finding the best fit
between the theoretical and the observed flux distribution of bursts
(unnormalized; i.e. using only the shapes of the distributions).  In
this fitting procedure, we fix the jet angle to be $\theta_{\rm
jet}=\pi/2$ (but consider the alternative $\theta_{\rm jet}=0.6$ rad,
the lowest value directly inferred from observations so far; see
below).  In practice, a distribution of viewing angles is inferred for
the theoretical GRBs (assuming random jet orientations), and compared
to the distribution $dn/d\theta$ inferred from the bursts detected by
{\it BATSE} (see Perna et al. 2003 for details of the analysis).  The
proportionality constant $K$ is then separately determined by imposing
that the total number of bursts $\int_0^{\pi/2}d\theta (dn/d\theta)$
above the BATSE sensitivity ($\sim 0.25~{\rm ph~sec^{-1}~cm^{-2}}$) be
667 yr$^{-1}$.  Once these two model parameters are determined, the
number of bursts observed at redshifts greater than $z$, over a time
interval, $\Delta t_{\rm obs}$, and solid angle, $\Delta \Omega$,
\begin{equation}
\label{eq:N_obs}
N(>z) = \frac{\Delta \Omega}{4 \pi} \Delta t_{\rm obs} \int_z^\infty d
z'\frac{\dot{\rho}_{\rm GRB}(z')}{(1+z')}\; \frac{dV(z')}{dz'}
\int_0^{\theta_{\rm max}(z')} P(\theta)d\theta,
\end{equation}
can be found for the {\it Swift} sensitivity ($F_{\rm lim}\sim
0.04~{\rm ph~sec^{-1}~cm^{-2}}$).  We adopt a solid angle
$\Delta\Omega=0.34$ sr, for the fully coded detector area of {\it
  Swift}.  In the above equation $P(\theta)d\theta = \sin\theta
d\theta$ is the probability of viewing a randomly oriented GRB at an
angle from the jet axis between $\theta$ and $\theta+d\theta$, the
factor $1/(1+z)$ accounts for time dilation, and $dV(z)/dz$ is the
comoving volume in our past light cone per unit redshift,
\begin{equation}
\label{eq:dVdz}
\frac{dV(z)}{dz} = 4 \pi c \frac{d_L^2(z)}{1+z} \left| \frac{dt}{dz}
\right|,
\end{equation}
where $c$ is the speed of light, and $d_L(z)$ is the luminosity
distance. The maximum angle $\theta_{\rm max}$ at which a burst can be
detected at a given redshift is found by numerically inverting the
equation
\begin{equation}
F_{\rm lim}=\frac{L(\theta_{\rm
max})}{4\pi d_L^2(z)(1+z)^\alpha},
\end{equation}
where $\alpha$ is the power--law spectral index of the photon number
spectrum of the burst (see Lamb \& Reichart 2000 for details).  

For illustration, in Figure~\ref{fig:fractions}, we show the fraction
of all bursts at redshift $z$ that would be detectable by {\it BATSE}
(dashed curves) and by {\it Swift} (solid curves).  The lower/upper
curves correspond to jet angles of $\theta_{\rm jet}=\pi/2$ and 0.6
rad, respectively.  In our fiducial model, {\it Swift} detects all
GRBs out to $z\sim 1$, and $\sim 1\%$ of GRBs at $z\sim 10$.

The value of the normalization constant, $K$, inferred from the {\it BATSE} data, is dependent
on the value of $L_{\rm min}$.  For example, in the CDM models with
instantaneous star--formation and $T_{\rm vir} > 10^4$ K, for our
fiducial choice of $\theta_{\rm jet}=\pi/2$, we find the best--fitting
value of $L_{\rm min}=1.5\times 10^{55}$ ph/sec, and the local GRB
rate of $\dot{\rho}_{\rm GRB}(z=0)=6~{\rm Gpc^{-3}~yr^{-1}}$, while
for $\theta_{\rm jet}=0.6$ rad, we find instead $L_{\rm min}= 10^{56}$
ph/sec, and $\dot{\rho}_{\rm GRB}(z=0)=1~{\rm Gpc^{-3}~yr^{-1}}$.

Nevertheless, we find that our subsequent predictions for the number
of bursts at high redshift by {\it Swift} are only mildly dependent on
the choice of $\theta_{\rm jet}$.  This is because as $L_{\rm min}$ is
increased, the normalization factor $K$ decreases. This tends to
compensate for the increase in the observable number of bursts due to
the fact that there is a larger fraction of bursts at higher
luminosities. As can be seen in Figure~\ref{fig:fractions}, at high
redshift only the bright power--law tail of GRBs are detectable, and
the two effects nearly cancel.  We note that the slope of $N(>z)$
depends on the power--law slope of the LF. Our model has $dn/dL
\propto L^{-\gamma}$ with $\gamma=2$ in the high luminosity tail --
for comparison, by fitting a single power--law to $dn/dL$, Firmani et
al. (2004) find a shallower slope with $\gamma\sim 1.6$.  If this
slope is accurate for high luminosities, it would somewhat increase
the number of high--$z$ GRBs.  Of course, once {\it Swift} data has
been gathered, it will provide new and independent constraints on the
GRB luminosity function, facilitating more accurate estimates of the
high--$z$ GRB rate.

\subsection{Uncertainties in GRB Redshift Evolution}
\label{sec:uncertain}

Our most important model assumptions are that $\epsilon_\ast$ and $K$
are constant.  
Observations of nearby dwarf galaxies (Taylor et al. 1999; Walter et
al. 2001) yield a range of efficiencies $\epsilon_\ast \approx$ 0.02
-- 0.08.  These values correspond to the low halo mass scales which
form at high redshifts, although there is no clear direct
correspondence between the first generation halos and these local
dwarfs.  Numerical simulations of metal--free star--formation at high
redshift (Abel, Bryan, \& Norman 2002; Bromm, Coppi \& Larson 2002)
suggest that the first generation of stars form in minihalos with
even lower efficiencies, $\epsilon_\ast \lsim$ 0.01. Such a reduction
in minihalo star--formation ($T_{\rm vir} < 10^4$ K) suggests
that at high redshifts, the true GRB rate is closer to our $T_{\rm
vir} > 10^4$ K curve.

On the other hand, since GRB progenitors involve high-mass stars
(MacFadyen \& Woosley 1999), and the stellar initial mass function
(IMF) is expected to be more top-heavy in the early universe, $K$
might increase at high redshifts, increasing $\GRBrd$.   We note
that the minimum mass of stars that loose their hydrogen envelope
and may thus produce GRBs can also strongly depend on metalicity.
Heger et al. (2003) find that this minimum mass increases from
$\approx 30 {\rm M_\odot}$ at solar metalicity to $\approx 100 {\rm
M_\odot}$ at zero metalicity (see their fig. 3).  However, the
trend is still more likely to be for $K$ to increase toward high
redshift, because $>30 {\rm M_\odot}$ stars are relatively rare
in local galaxies with a Salpeter IMF, while the simulations of
Abel et al. (2002) and Bromm et al. (2002) suggest that
metal--free star formation at high redshift may produce exclusively
$>100 {\rm M_\odot}$ stars.  
Thus, since they are expected to act in opposite directions, the
redshift evolutions of $\epsilon_\ast$ and $K$ might compensate
somewhat for each other.  The rather high angular momentum of the
collapsing star required to produce a GRB is more easily achieved if
the star is in a binary, and binaries are found to be more frequent at
low metallicity (i.e. high $z$; Fryer et al. 1999).  It is, however
unclear if binaries do form in the first generation of truly
metal--free halos.
Simulations following direct cosmological initial conditions by Abel,
Bryan \& Norman (2002) find a single star with no further
fragmentation; however, different simulations with somewhat more artificial initial conditions (a
rotating cylinder), find efficient
binary--formation (Saigo et al. 2004).

It is conceivable that at high redshifts, a WDM universe could mimic
the GRB rates of a CDM universe, by compensating for a loss of
small--scale power with higher efficiencies of GRB production (e.g. a
higher $K$ value at high redshifts), and visa--versa.  However, we
note that such a redshift evolution is unlikely to be sharp enough to
significantly impact our conclusions.  Even a power--law evolution of
$K$ would be insufficient to compensate for the exponential
suppression of small--scale power in our models.  For example, from
Figure \ref{fig:grbvsz} we note that in order for GRB rates in WDM
models with $m_x \approx 2$ keV to match GRB rates in CDM models at
$z>10$, 
the product $\epsilon_\ast K$ would need to be a factor of 10 higher
than the overall average value determined from the lower--redshift
{\it BATSE} sample.  For $m_x \approx 1$ keV, $\epsilon_\ast K$ would have to be
$\sim 100$ times larger at $z>10$. These differences only increase
with increasing redshift, and if such an increase is present, it can,
in principle, be detected by studying the {\it shape} of the GRB
redshift distribution function (see \S~\ref{sec:fract_constraints}).

Recently, Wise \& Abel (2004) calculated primordial supernovae (SNe)
rates with a semi-analytic analysis of feedback mechanisms and
evolution of primordial stellar environments, constraining their
results with the measured {\it WMAP} optical depth to electron
scattering, $\tau_e = 0.17$.  
With ab--initio knowledge of GRB progenitors, an analogous analysis
could be performed to predict the GRB rate as a function of redshift,
replacing the proportionality constant, $K$, assumed in this paper.
In practice, our current knowledge of the physics of GRB progenitors
and of the various feedback processes in the early universe is highly
uncertain, and such an approach would introduce additional free
parameters.  Nevertheless, this approach could be useful in the
future, since some of the free parameters may be independently
constrained. For example, in addition to GRBs, early star--formation will be accompanied by observable SN explosions (e.g. Miralda-Escud\'e \& Rees 1997), the production of heavy elements
(e.g. Haiman \& Loeb 1997), remnant stellar black holes
(e.g. Volonteri et al. 2003; Volonteri \& Perna 2005), and the reionization of the IGM.  Observations of these effects can further reduce the uncertainties associated with
early star--formation history and GRB progenitors, and thus
ultimately reduce uncertainties on the expected primordial GRB event rates (for a
discussion of such future constraints, see Wise \& Abel 2004, who also
emphasize that SNe observations can constrain many properties, such as
mass, luminosity, metallicity, and redshift, of the
progenitors). Finally, early ``mini--galaxies'' may also be directly
detectable (Haiman \& Loeb 1997) by the {\it James Webb Space
Telescope (JWST)}.\footnote{See www.jwst.nasa.gov}

\section{Constraints on Structure Formation Models}
\label{sec:constraints}

The model outlined in the previous sections can be used to compute the
evolution of the GRB rate with redshift, as well as the flux
distribution of the bursts, allowing us to incorporate the detection
threshold of {\it Swift}. In this section, we present constraints on
WDM models first from the total number of GRBs, and then a potentially
less model--dependent constraint from the distribution of a {\it
luminosity--limited} sub--sample of the bursts.

\vspace{+0\baselineskip} \myputfigure{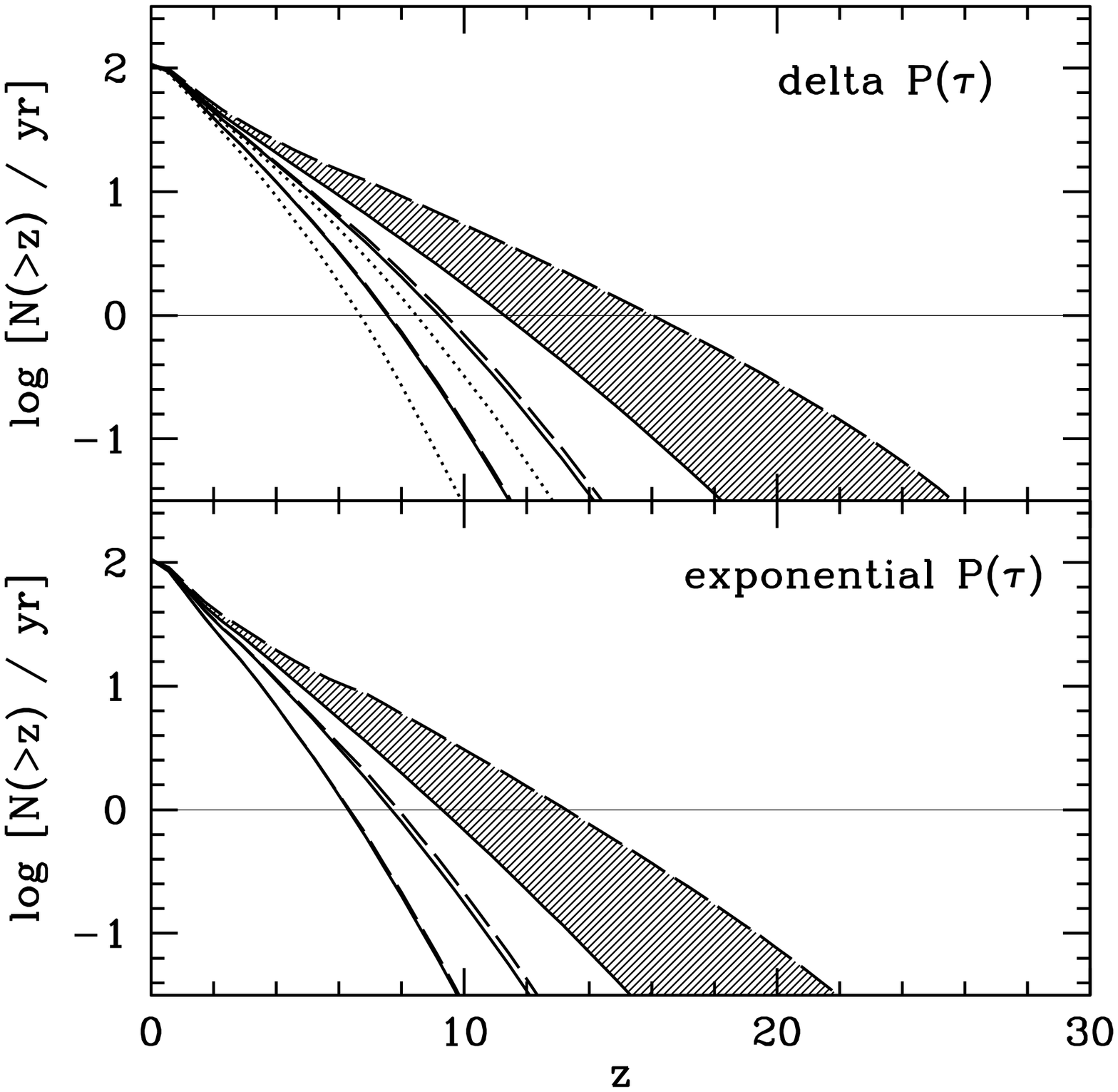}{3.3}{0.5}{.}{0.}
\vspace{-1\baselineskip} \figcaption{Expected {\it Swift} detection
rates of GRBs occurring at redshifts greater than $z$.  The curves
correspond to the same models as shown in
Figure~\ref{fig:sfrd_fract}, and the horizontal line denotes the detection
threshold of 1 GRB/yr.  The {\it top panel} assumes instantaneous
star--formation; the {\it bottom panel} assumes a finite exponential
spread in star--formation times. \label{fig:grbvsz}}
\vspace{+1\baselineskip}

\subsection{Absolute GRB Detection Rates}
\label{sec:abs_constraints}

The most straightforward constraints follow from the total {\it Swift}
burst detection rates.  In Figure~\ref{fig:grbvsz}, we show the
cumulative {\it Swift} GRB detection rates produced by our models.
The curves correspond to the same models as shown in
Figure~\ref{fig:sfrd_fract} above, with the shaded area enclosing the
range of expected distributions in CDM models for different limiting
virial temperatures for star--formation in early halos. The top panel
assumes instantaneous star--formation, and the bottom panel assumes a
spread of star--formation times as discussed above.  The horizontal
line denotes a detection rate of 1 GRB per year, roughly the lower
limit for detections in {\it Swift}'s two--year nominal operation.

Figure~\ref{fig:grbvsz} shows that detections of bursts originating at
redshifts larger than $z \sim 15$ are improbable even in CDM models.
Constraints on WDM models can be simply read off this figure by noting
the redshifts at which the curves corresponding to a given model
intersect the horizontal line.  For example, for a WDM particle mass
of $m_x$ = 1 keV, a detection rate of one $z>7$ GRB per year is
expected. That number drops to one $z>8$ GRB per 10 years, which is
unlikely to be detectable.  {\it A detection of a single $z \gsim
10$ GRB by {\it Swift} would rule out models incorporating WDM or
equivalent power spectrum cut--offs of $m_x \lsim 2$ keV ($R_c
\gsim 0.087$ Mpc), and would therefore significantly improve the
existing constraints.}

These constraints on models with an exponential cut--off in the
small--scale power spectrum are summarized further in
Table~\ref{tab:summary}.  The low end of the quoted redshift range in
each case in the top five rows corresponds to the exponential
star--formation and the high end to instantaneous star--formation (the
models that include the effective pressure of WDM particles, listed in
the bottom three rows, show only the latter).  The right--most column
in the table shows the redshifts beyond which 10 GRBs per year would
be detectable.

The constraints shown in Table~\ref{tab:summary} have been derived
assuming a fixed cosmological model, varying only the cut--off scale
for the power spectrum. However, in principle, the effects of the
cut--off can be mitigated by changing other parameters.  In particular,
the GRB rate depends exponentially on the normalization of the
power--spectrum $\sigma_8$.  In order to quantify the robustness of
our results, we have re--computed the redshifts corresponding to 1 and
10 GRBs/yr in the CDM model with $T_{\rm vir} > 10^4$ K and in the model
with the power--spectrum cutoff at $R_c=0.193$ Mpc, but using a high
value of $\sigma_8=1$ rather than our fiducial choice of
$\sigma_8=0.85$.  We have found that this increases the redshifts by
$\Delta z < 2$; in particular, the $R_c=0.193$ Mpc model could still
be ruled out by observing 1 GRB/yr at $z\gsim 9.7$.

\begin{table}[ht]
\caption{The redshift, in several WDM models, beyond which the Swift
satellite would detect 1 or 10 GRBs/year.}
\vspace{-0.8cm}
\label{tab:summary}
\begin{center}
\begin{tabular}{lll}
\tablewidth{3in}\\
\hline
\hline
Model  & 1 GRB/yr & 10 GRBs/yr\\
\hline
\hline
CDM ($T_{\rm vir}>10^4$K)  & $z \gsim$ 9.40 -- 11.6  & $z \gsim$ 4.8 -- 5.9\\
CDM ($T_{\rm vir}>300$K)  & $z \gsim$ 13.4 -- 16.0  & $z \gsim$ 6.4 -- 7.6\\
\hline
$P(k)$ suppr ($R_c \lsim$ 0.193Mpc)  & $z \gsim$ 6.3 -- 7.5 & $z \gsim$ 3.5 -- 4.3\\
$P(k)$ suppr ($R_c \lsim$ 0.087Mpc)  & $z \gsim$ 7.8 -- 9.3 & $z \gsim$ 4.2 -- 5.1\\
$P(k)$ suppr ($R_c \lsim$ 0.055Mpc)  & $z \gsim$ 8.4 -- 10 & $z \gsim$ 4.5 -- 5.4\\
\hline
WDM ($m_x \gsim$ 1keV) & $z \gsim$ 6.7 & $z \gsim$ 3.9\\
WDM ($m_x \gsim$ 2keV) & $z \gsim$ 8.5 & $z \gsim$ 4.8\\
WDM ($m_x \gsim$ 3keV) & $z \gsim$ 9.5 & $z \gsim$ 5.2\\
\hline
\hline
\end{tabular}\\
\end{center}
\end{table}

The corresponding constraints in models with a red tilt or a running
of the scalar index are summarized in Table~\ref{tab:tiltsummary}.
Overall, in the CDM model with the more stringent limit of $T_{\rm
vir} > 10^4$ K for star--formation, the effect of introducing a tilt with
$n_s=0.9$, or a running of the index with $\alpha=dn_s/d\ln k=-0.05$
is comparable to that of the exponential cut--off in the power spectrum
due to a $m_x \approx 2$ keV WDM particle: the GRB rate falls below
1/yr at $z\gsim 8$.  However, as the table shows, these constraints
are less robust than those for WDM. The suppression of the power
spectrum is a shallow function of scale, and, unlike in the case of
the exponential suppression, the GRB rate can be increased by either a
relatively modest change in the normalization $\sigma_8$, or by
allowing star--formation in smaller halos down to $T_{\rm vir}=300$ K.
The table suggests that a detection rate of 1 GRB/yr even at $z\gsim
14$ cannot be used to distinguish a tilt of $n_s\approx 0.9$ from
models with $n_s=1$ but with a lower $T_{\rm vir}>300$ K and a higher
$\sigma_8$.  However, a detection rate of 1 GRB/yr at $z\gsim 10$
would lead to a relatively robust upper limit on the running of the
spectral index, $\alpha\gsim -0.05$.

\begin{table}[ht]
\caption{The redshift in models with a red--tilted or running scalar index,
 fixed Swift GRB rates, as in Table~\ref{tab:summary}.}
\vspace{-0.8cm}
\label{tab:tiltsummary}
\begin{center}
\begin{tabular}{lll}
\tablewidth{3in}\\
\hline
\hline
Model  & 1 GRB/yr & 10 GRBs/yr\\
\hline
\hline
\multicolumn{3}{c}{Red Tilt ($n_s=0.9$)}\\
\hline
$T_{\rm vir}>10^4$K  & $z \gsim$ 7.7 -- 9.3  & $z \gsim$ 4.1 -- 5.0\\
$T_{\rm vir}>300$K  & $z \gsim$ 10.5 -- 12.6  & $z \gsim$ 4.8 -- 6.3\\
$\sigma_8=1$ ($T_{\rm vir}>10^4$K)  & $z \gsim$ 9.1 -- 11.1  & $z \gsim$ 4.8 -- 5.8\\
$\sigma_8=1$ ($T_{\rm vir}>300$K)  & $z \gsim$ 12.2 -- 14.7  & $z \gsim$ 5.8 -- 7.3\\
\hline
\multicolumn{3}{c}{Running ($\alpha=-0.05$)}\\
\hline
$T_{\rm vir}>10^4$K  & $z \gsim$ 7.4 -- 8.9  & $z \gsim$ 4.0 -- 4.9\\
$T_{\rm vir}>300$K  & $z \gsim$ 8.8 -- 10.5  & $z \gsim$ 4.3 -- 5.4\\
$\sigma_8=1$ ($T_{\rm vir}>10^4$K)  & $z \gsim$ 8.7 -- 10.5  & $z \gsim$ 4.6 -- 5.6\\
$\sigma_8=1$ ($T_{\rm vir}>300$K)  & $z \gsim$ 10.2 -- 12.3  & $z \gsim$ 5.1 -- 6.4\\
\hline
\hline
\end{tabular}\\
\end{center}
\end{table}

\subsection{Redshift Distribution of a Luminosity--Limited Sample}
\label{sec:fract_constraints}

A method that is less dependent on models of the GRB LF, and that could, in
principle, discriminate against WDM models and models with a power
spectrum cut--off is to construct a redshift distribution of a {\it
luminosity--limited} sub--sample of the observed bursts.  Under the
relatively weak assumption that the LF of GRBs does not evolve with
redshift (but without any other assumptions about the LF), this
redshift distribution would be proportional to the distribution of all
bursts.  In the bottom panels of Figure~(\ref{fig:sfrd_fract}), we
plot the fractional distribution of all bursts,
\begin{equation}
f \equiv \int_z^\infty dz' \frac{\dot{\rho}_{\rm GRB}(z')}{(1+z')}
\frac{dV(z')}{dz'} \times \left[\int_0^\infty dz' \frac{\dot{\rho}_{\rm
GRB}(z')}{(1+z')} \frac{dV(z')}{dz'}\right]^{-1}.
\label{eq:allGRBs}
\end{equation}
This quantity is then independent of the value of the normalization
$K$ and of the shape of the GRB LF discussed above.

Such intrinsic event rate distributions can be constructed by tracking
the redshift distribution only for the subset of GRBs above a minimum
luminosity that allows one to see them up to a given high redshift,
and neglecting GRBs which are fainter than this threshold.  By
construction, all of the remaining GRBs are luminous enough to be
detectable up to the chosen redshift and so trace out the intrinsic
event rates, $f$ (i.e. independently of the shape of the LF and $K$).
The high sensitivity of {\it Swift} makes this still a sizable
sample. For example, in the case of the standard CDM model, we find
that $\approx 30\%$ of the bursts would still remain by choosing only
the ones above the minimum luminosity that allows for their detection
up to $z\sim 10$.  Likewise, $\approx 20\%$ of bursts would constitute
the sample with the minimum luminosity that allows for their detection
up to $z\sim 15$.  The bottom panels of Figure~\ref{fig:sfrd_fract}
reveal that $\sim 50\%$ of these remaining bright GRBs would be at
redshifts $z \gsim 5$, where the curves start showing significant
deviations from one another.  In the CDM case, this would translate to
a sample of $\sim 20$ high--redshift ($z \gsim 5$) GRBs with which to
trace out the redshift distribution in the bottom panels of
Figure~\ref{fig:sfrd_fract}.

Finally, we note that even for this high tail of the LF, the
efficiency with which ultra--high redshift GRBs can be identified is
likely to be a decreasing function of redshift itself. Although the
network of follow--up instruments throughout the globe, combined with
the aggressive search of high-$z$ GRBs, is likely to partly compensate
for the difficulties of identifying the GRBs at the highest redshifts,
such selection effects would still have to be carefully folded in to
an analysis of the real data (see, e.g., Gou et al. 2004 for a recent
discussion of the detectability of the afterglows of high--redshift
GRBs).

\section{Conclusions}
\label{sec:conclusions}

We find that high--redshift GRB detections are effective at
constraining the small--scale power spectrum of cosmological density
fluctuations.  Assuming that GRBs trace out the cosmic star--formation
rate, we generate expected GRB detection rates for various
cosmological models which include suppression of the density
fluctuations on small--scales.  The effects of such suppressions
become more notable at high redshifts, where the characteristic
collapse scales are smaller.  Correspondingly, we are able to obtain
strong constrains from high--redshift GRB detections.  For example, a
{\it Swift} detection of a single GRB at $z\gsim 10$, or 10 GRBs at
$z\gsim5$, would rule out an exponential suppression of the power
spectrum on scales below $R_c=0.09$ Mpc (exemplified by warm dark
matter models with a particle mass of $m_x=2$ keV).  Constructing the
intrinsic fractional GRB distribution from a luminosity--limited sample would
provide additional constraints that are independent from the
uncertainties in the normalization between GRB and star--formation
rates, and from the uncertainties in the intrinsic luminosity function of GRBs.  We find that
a detection of 1 GRB/yr at $z \gsim 12$ also provides an upper limit
on the running of the spectral index, $\alpha \gsim -0.05$, thus
potentially placing constraints on the inflationary potential.

\acknowledgments{
We thank Tom Abel and Avi Loeb for useful comments. This work was
supported in part by NSF through grants AST-0307200 and AST-0307291
(to ZH) and by NASA through grants NAG5-26029 and SWIF03-0000-0033 (to
ZH) and SWIF03-0020-0058 (to RP).}

\end{document}